\documentclass{aa}
\usepackage{graphicx, natbib,epsfig,amsmath,amssymb, mathrsfs, txfonts}
\DeclareGraphicsExtensions{.eps, .jpg, .ps}
\bibpunct{(}{)}{;}{a}{}{,}
\begin{document}

\title{Speckle temporal stability in XAO coronagraphic images}
\author{P.\ Martinez\inst{1} \and C.\ Loose\inst{2} \and  E.\ Aller Carpentier\inst{2}   \and M.\ Kasper\inst{2}}
\institute{UJF-Grenoble 1/CNRS-INSU, Institut de Plan\'{e}tologie et d'Astrophysique de Grenoble UMR 5274, Grenoble, F-38041, France
\and  European Southern Observatory, Karl-Schwarzschild-Stra\ss{}e 2, D-85748, Garching, Germany} 
\offprints{patrice.martinez@obs.ujf-grenoble.fr}

\abstract
{The major noise source limiting high-contrast imaging is due to the presence of quasi-static speckles. Speckle noise originates from wavefront errors caused by various independent sources, and it evolves on different timescales pending to their nature. An understanding of quasi-static speckles originating from instrumental errors is paramount for the search of faint stellar companions. Instrumental speckles average to a fixed pattern, which can be calibrated to a certain extent, but their temporal evolution ultimately limit this possibility.
}
{This study focuses on the laboratory evidence and characterization of the quasi-static pinned speckle phenomenon. Specifically, we examine the coherent amplification of the 
static speckle contribution to the noise variance in the scientific image, through its interaction with quasi-static speckles. 
}
{The analysis of a time series of adaptively corrected, coronagraphic images recorded in the laboratory enables the characterization of the temporal stability of the residual speckle pattern in both direct and differential coronagraphic images.} 
{We estimate 
that spoiled 
and fast-evolving quasi-static speckles present in the system at the angstrom/nanometer level are affecting the stability of the static speckle noise in the final image after the coronagraph. 
The temporal evolution of the quasi-static wavefront error exhibits linear power law, which can be used in first order to model quasi-static speckle evolution in high-contrast imaging instruments.}
{}

\titlerunning{Speckle temporal stability in XAO coronagraphic images}
\maketitle

\section{Introduction}
Direct detection of planets around nearby stars is challenging due to the required contrast and angular separation of the companion from its bright parent star.
From the ground, the first issue to overcome is the large, dynamical wavefront error that the atmosphere generates. 
Real-time adaptive optics systems measure and correct these down to fundamental limitations 
(fitting error, spatial aliasing, wavefront sensor noise, and servo-lag error).
The next generation of high-contrast imaging instruments such as SPHERE \citep{SPHERE}, GPI \citep{GPI}, HiCIAO \citep{HiCIAO}, and Project 1640 \citep{P1640} will improve the actual level of correction by reducing these fundamentals. High-order deformable mirrors, combined with more sensitive wavefront sensors, i.e.\ extreme adaptive optics (XAO) systems, will enable a dramatic reduction of the fast-varying speckle noise floor left uncorrected upon the actual performance.

For an ideal coronagraphic 
system, the stellar residual noise would average out over time. Owing to the random nature of this speckle noise, it leaves a smooth halo for which the intensity variance converges to null contribution for an infinitely long exposure. In such conditions, the fundamental noise floor in the science image is set by the photon noise. 
However, ideal instruments do not exist, and wavefront aberrations introduced in the optical train 
contribute to the noise budget. 
In contrast to the random and uncorrelated noise suppressed through averaging, static and quasi-static speckles do affect the sensitivity of an observing sequence. 
For instance, optical quality (polishing) or misalignment errors generate static speckles that constitute a deterministic contribution to the noise variance, and thereby 
long-timescale wavefront errors. Though it corresponds to an additional perturbation, a deterministic contribution can easily be calibrated.
Nonetheless, observing acquisitions have shown that the major noise source limitation in high-contrast imaging is due to the presence of another class of instrumental speckles, the \textit{quasi-static} speckles \citep{Beuzit97, Oppenheimer01, Marois03, Boccaletti03, Boccaletti04, Hinkley07}.  
Quasi-static speckles correspond to slowly-varying instrumental wavefront aberrations, amplitude and phase errors, present in the system, which finally dominate the companion signal. Such speckles originate from various causes, among others mechanical or thermal deformations, or Fresnel effects \citep[highly chromatic speckles originating from phase-to-amplitude conversion due to out-of-pupil plane optics,][]{Marois06}, and evolve on a shorter timescale than long-lived aberrations. \\
The temporal evolution of these quasi-static speckles is needed for the quantification of the gain expected with angular differential imaging \citep[ADI,][]{Marois2006}, as well as to determine the timescale on which speckle nulling techniques should be carried out. 
Several authors have investigated the decorrelation timescale of quasi-static residuals in the particular context of ADI but at moderate 20-40~$\%$ Strehl levels \citep{Marois2006, 2007ApJ...660..770L}, though to our knowledge, in the context of very high Strehl (XAO systems), no information is present in the literature.
Understanding of these time-variable quasi-static speckles, and especially their interaction with other aberrations, referred to as the \textit{pinning effect}, is paramount for the search of faint stellar companions.
The timescale of quasi-static speckles evolution is essential to understand and predict the performance of XAO instruments.

The propagation through a coronagraph in the presence of aberrations, and the description of the pinned speckle phenomenon have been intensively analyzed in \citet{Soummer07} from a statistical and numerical point of view. 
By adopting the same formalism we can write down the wavefront complex amplitude at the entrance pupil as the coherent sum of four terms,
\begin{equation}
\Psi (r) = [A + A_{s}(r) + a_{1}(r) + a_{2}(r)] \times P(r).
\label{champ}
\end{equation}
\noindent Here, the function $P(r)$ describes the normalized telescope transmission, $A$ is a deterministic term corresponding to a perfect, i.e.\ plane wave, $A_{s}(r)$ is a deterministic, complex term corresponding to the static aberrations, $a_{1}(r)$ is the atmospheric part of the wave front (a random term with zero mean), $a_{2}(r)$ is the contribution of the quasi-static speckle part of the wavefront, and $r$ is the coordinate vector. In addition, $|A^2|$ represents the Strehl ratio $S$ of the system. It is assumed that the lifetimes, $\tau_1$ of $a_1(r)$ and $\tau_2$ of $a_2(r)$, are such that $\tau_2 > \tau_1$. In this context, $N_1$ and $N_2$ describe the number of speckle realizations associated with each process during an exposure of duration $\Delta t$, and finally, $N$, the ratio of $N_1$ over $N_2$, is the number of fast speckle realization during a slow-speckle lifetime $\tau_2$/$\tau_1$. 

\citet{Soummer07} proposed the following analytical expression for the variance of the intensity, including speckle and photon noise in the presence of static, quasi-static, and fast varying aberrations,
\begin{equation}
\sigma^{2}_I= N \left( I^{2}_{s1} + NI^{2}_{s2} + 2I_c I_{s1} + 2N I_c I_{s2} + 2 I_{s1}I_{s2} \right) + \sigma^{2}_p,
\label{variance}
\end{equation}
\noindent where $I$ denotes the intensity, and $\sigma^{2}_p$ is the variance of the photon noise. The intensity produced by the deterministic part of the wavefront, including static aberrations, is denoted by $I_c$, while the $I_s$ terms correspond to the halo produced by random intensity variations, i.e.\ atmospheric and quasi-static contributions. 

Equation~\ref{variance} is used 
to describe the effect of quasi-static speckles and their interactions with other aberrations present in a system. In this generalized expression of the variance, several contributions can be identified by order of appearance: (1/) 
the atmospheric halo, (2/) the quasi-static halo, (3/) the atmospheric pinning term, where the static aberrations are amplified by the fast evolving atmospheric speckles, (4/) the speckle pinning of the static by quasi-static speckles, 
and finally (5/) the speckle pinning of the atmospheric speckles by quasi-static speckles. 
The noise budget of Eq.~\ref{variance} provides useful insights in 
the understanding of high-contrast imaging, especially in the presence of a coronagraph. It also enables a general comprehension of the several flavors of the pinning phenomenon. The interest of the combination of adaptive optics system, coronagraphy,  speckle calibration, active correction, speckle nulling, and post-processing methods to remove most of the terms contributing to the noise variance is enlightened 
here. 

The present study is concerned with the laboratory evidence of contribution (4/) to this noise budget in XAO coronagraphic images. We examine and quantify the speckle pinning of the static by the quasi-static speckles in a real system delivering highly corrected, coronagraphic images when no atmospheric contribution is present, i.e., when the contribution of $I_{s1}$ to Eq.~\ref{variance} can be neglected. In this situation, Eq.~\ref{variance} can be simplified such that:
\begin{equation}
\sigma^{2}_I \simeq \left(I^{2}_{s2} +  2 I_c I_{s2}  \right) + \sigma^{2}_p,
\label{variance2}
\end{equation}
\noindent and the present study focuses on the effect of the cross-term $I_c I_{s2}$. 
\noindent The paper reads as follows: In Sect.~2 we detail the test apparatus and data reduction aspects, while in Sect.~3 we present and discuss the results. In Sect.~4 we attempt to discussed our results in the broader context of actual XAO instruments. Finally, we draw conclusions in Sect.~5.

\section{Laboratory conditions}
\subsection{Experimental setup}
The High-Order Testbench \citep[HOT, ][]{HOTbench} is a versatile high-contrast imaging, adaptive optics bench developed at the European Southern Observatory (ESO) to test and optimize different techniques and technologies, in particular wavefront sensors, coronagraphs, speckle calibration methods, and image post-processing. It offers a unique opportunity to operate several critical components in a close-to-real instrument environment, by reproducing the conditions encountered at a telescope, e.g.\ the ESO Very Large Telescope (VLT).

HOT comprises a turbulence generator with phase screens to simulate real seeing conditions. The light sources are a broadband, superluminescent semiconductor diode (SLED) module from EXALOS, centered in the near-infrared at 1.6~$\mu$m, coupled with a halogen white light lamp for wavefront sensing in the visible. A mask reproducing the footprint of the VLT pupil is installed on a 60-element bimorph deformable mirror (DM) which is held by a tip-tilt mount and employed to correct for static aberrations. A 32\,$\times$\,32~actuator micro-DM, an electrostatic MEMS device, can be used to correct dynamical aberrations. A beam splitter diverts 50~\% of the light towards a wavefront sensor
, while the rest is transmitted towards the coronagraph and the infrared camera, a 1k\,$\times$\,1k HAWAII detector. The infrared camera offers various possibilities of near-IR filters, while during the experiment a narrow-band  $H$ filter was used ($\Delta \lambda / \lambda = 1.4~\%$).
All optics are mounted on an air-suspended table fully covered with protection panels and forming a nearly closed box, set in a dark room.
HOT features both a pyramid wavefront sensor (PWS) and the Shack-Hartmann Sensor (SHS) 
which was used for this study. 
The implemented coronagraph is an Apodized Pupil Lyot Coronagraph \citep[][APLC]{Soummer03}, with a focal plane mask diameter of 4.5~$\lambda/D$. This corresponds to an inner working angle (IWA) of 2.3~$\lambda/D$ (90~mas in $H$-band). The IWA offers a qualitative description of the minimum separation from the bright on-axis star, at which the coronagraphic design allows the detection of a faint companion.
The APLC combines pupil apodization with a hard-edged focal plane occulter and an undersized pupil-plane stop with 90~\% throughput. 
It is installed at an F/48 beam. 
Further details of the prototype can be found in \citet{microdots1}.

Because the goal of this study is to examine the interaction between quasi-static and static aberrations, the phase screens of the bench were replaced by mirrors. This setup avoids the presence of dynamical aberrations and justifies the approximation made in Eq.~\ref{variance2}. 
The experiment uses the micro-DM in combination with the bimorph DM to correct static aberrations. The latter is operated in open-loop, i.e.\ by statically applying a corrective voltage pattern in addition to the fix operating voltage which sets the mirror surface to its initial shape.

\begin{figure*}[ht!]
\centering
\includegraphics[width=5.6cm]{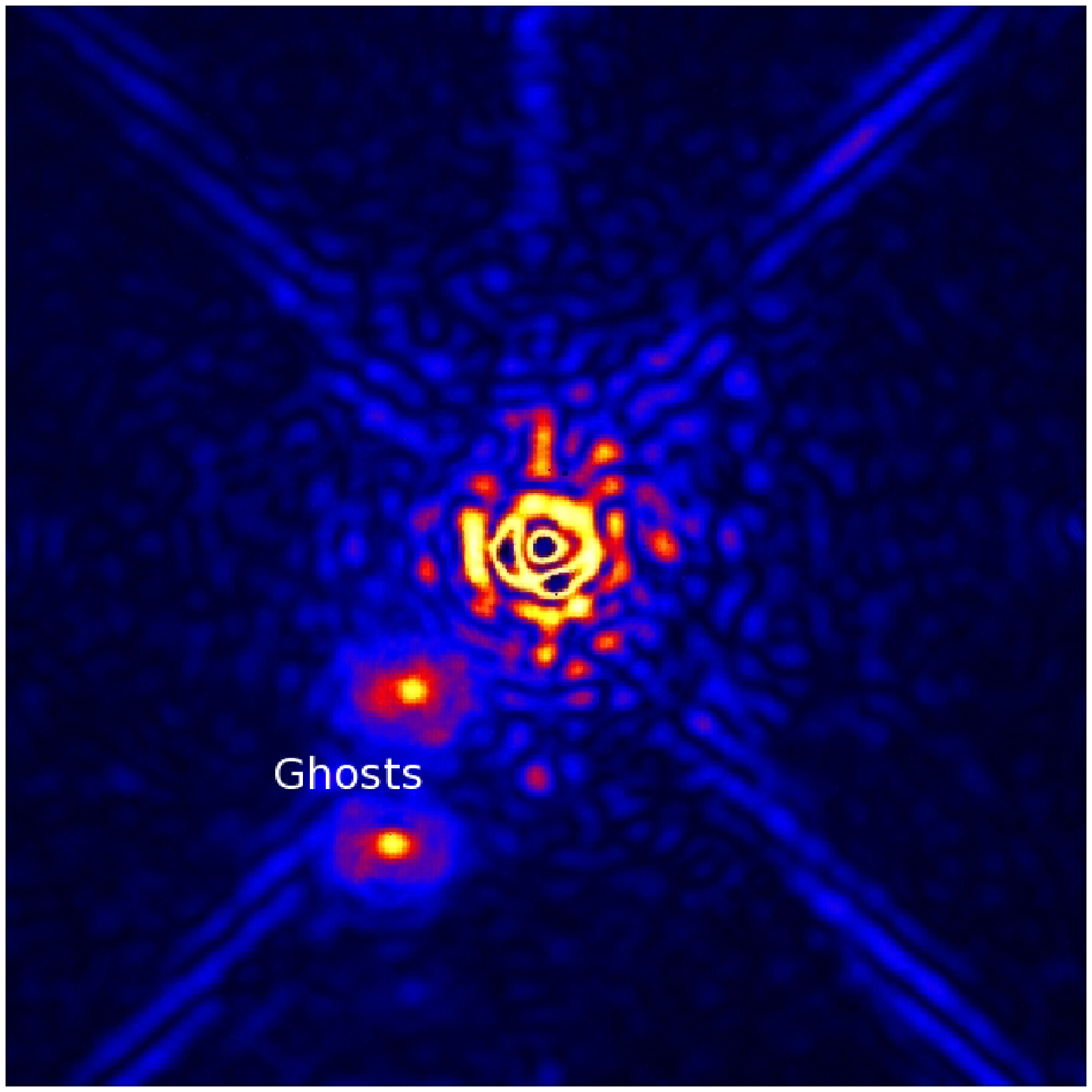}
\includegraphics[width=5.6cm]{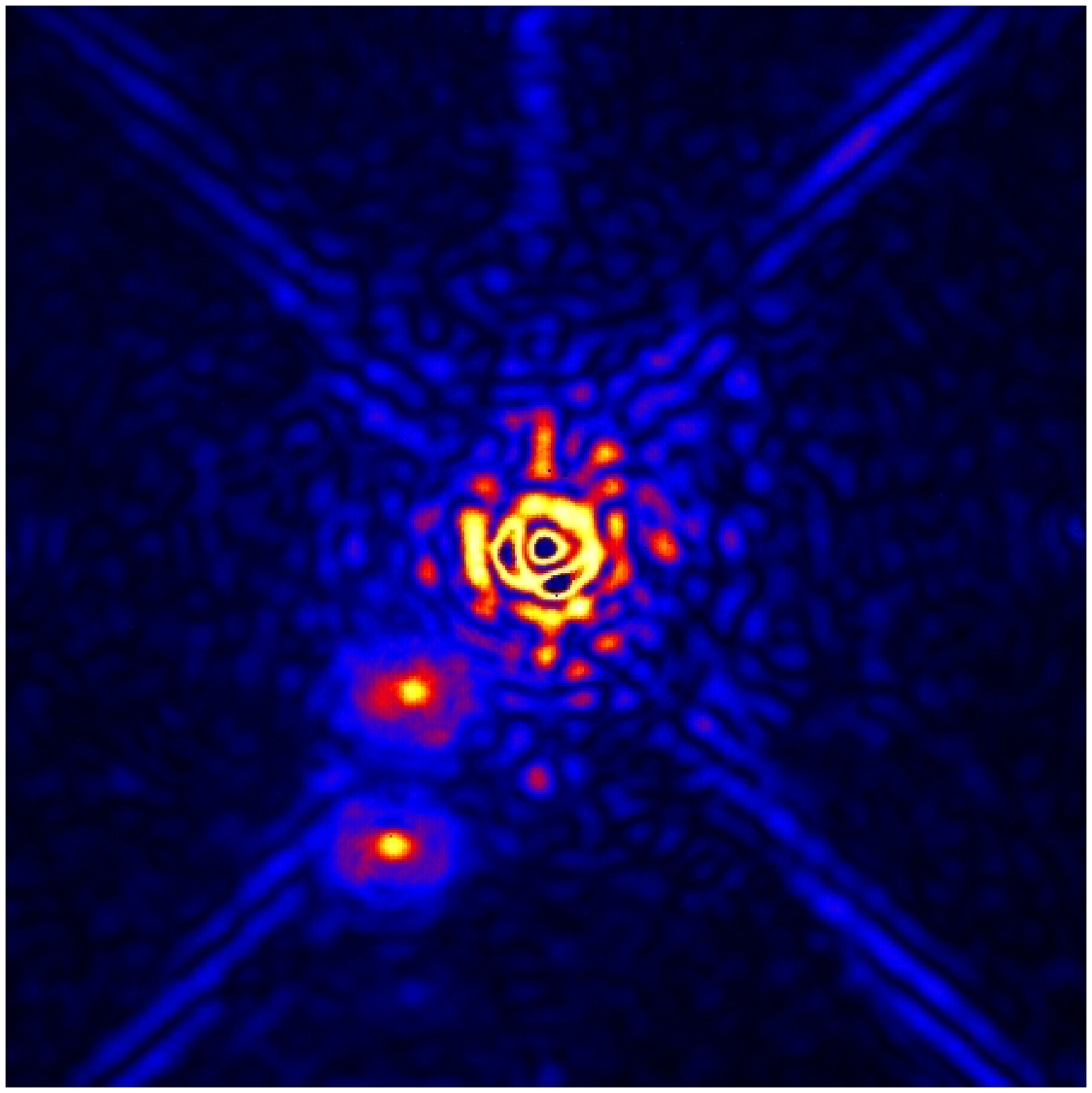}
\includegraphics[width=5.6cm]{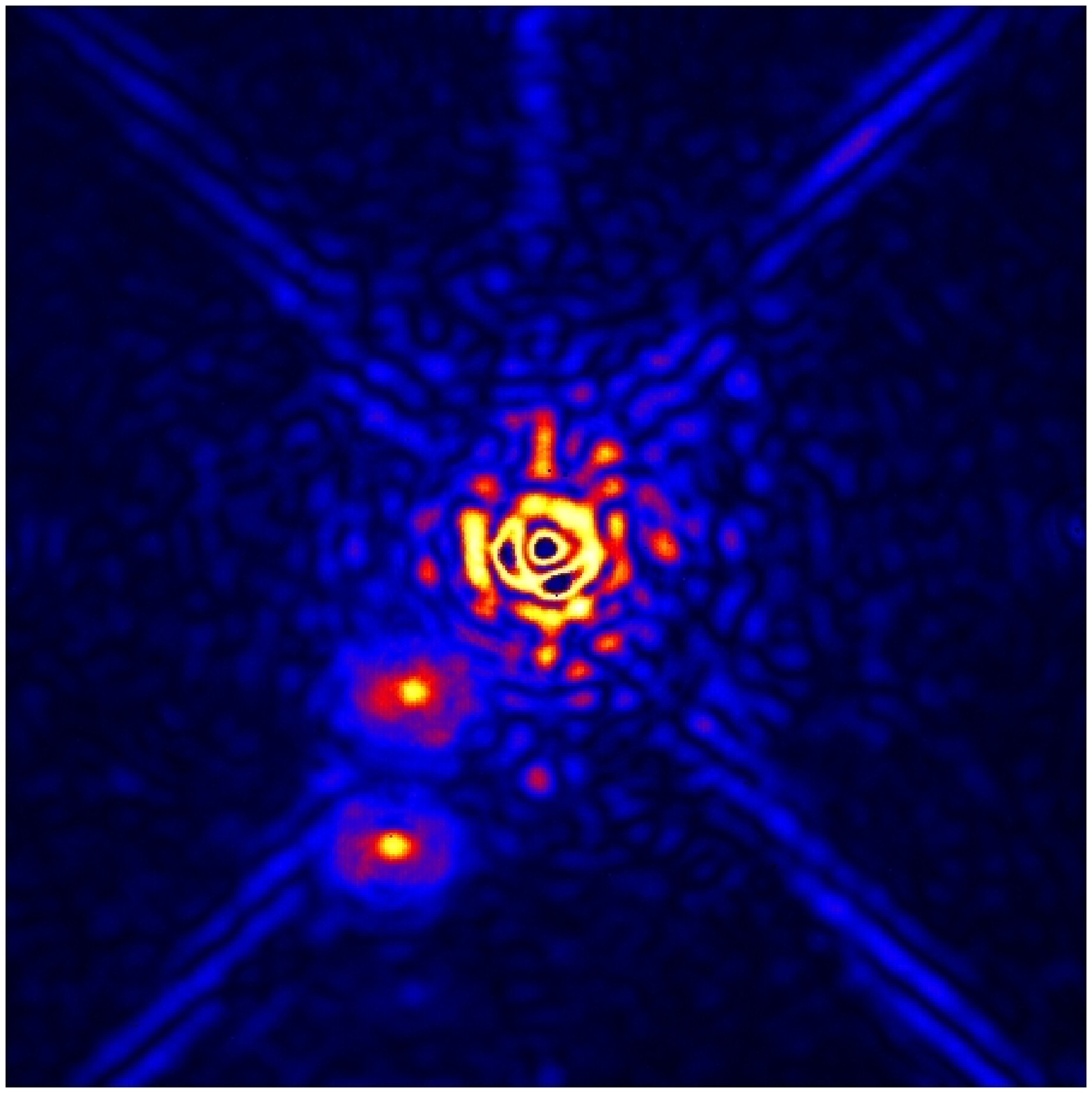}
\caption{Coronagraphic images recorded at $t_{0}$, $t_{0} + 3$~mn, and $t_{0} + 6$~mn. The signal is intentionally saturated in the image core to emphasise the speckle signal in the halo. The Strehl ratio is  $\sim$~92~\%. The arbitrary color scale and dynamic range (identical for the three images) were chosen to enhance the contrast for the sake of clarity. The presence of two ghosts is highlighted in the left figure.} 
\label{Images}
\end{figure*}

The SHS is used to drive the Tip-Tilt mount in closed-loop, countering the main contribution of bench turbulence.
The static aberrations which are common-path wavefront errors, are thereby reduced from $\sim$~200 to $\sim$~50 nm rms using the bimorph DM in open loop, i.e.\ by applying a corrective voltage pattern to the initial voltage pattern that set-up the mirror \citep{HOTbench}. The wavefront sensor non-common path wavefront errors have been measured using a reference fibre at the entrance of the sensor, and slope offsets have been applied to the DM not to take into account these aberrations in the correction. Additionally, some slope offsets have been applied to correct for non-common path errors from the near-IR optical path (the initial wavefront error level is estimated to 25 nm rms) on the basis of the Point Spread Function (PSF) image quality. Further details can be found in \citet{HOTbench, Martinez10, Martinez11}. Therefore, the static wavefront errors in the system can be estimated to be between 50 and 75~nm rms. In such conditions, the near-IR Strehl ratio is $\sim$~92~\%. Qualitative temporal analysis of a time series of coronagraphic images over two days have shown the constance of the residual speckle pattern; images recorded on the timescale of several minutes are presented in Fig.~\ref{Images}. 
\begin{figure*}[ht!]
\centering
\includegraphics[width=5.6cm]{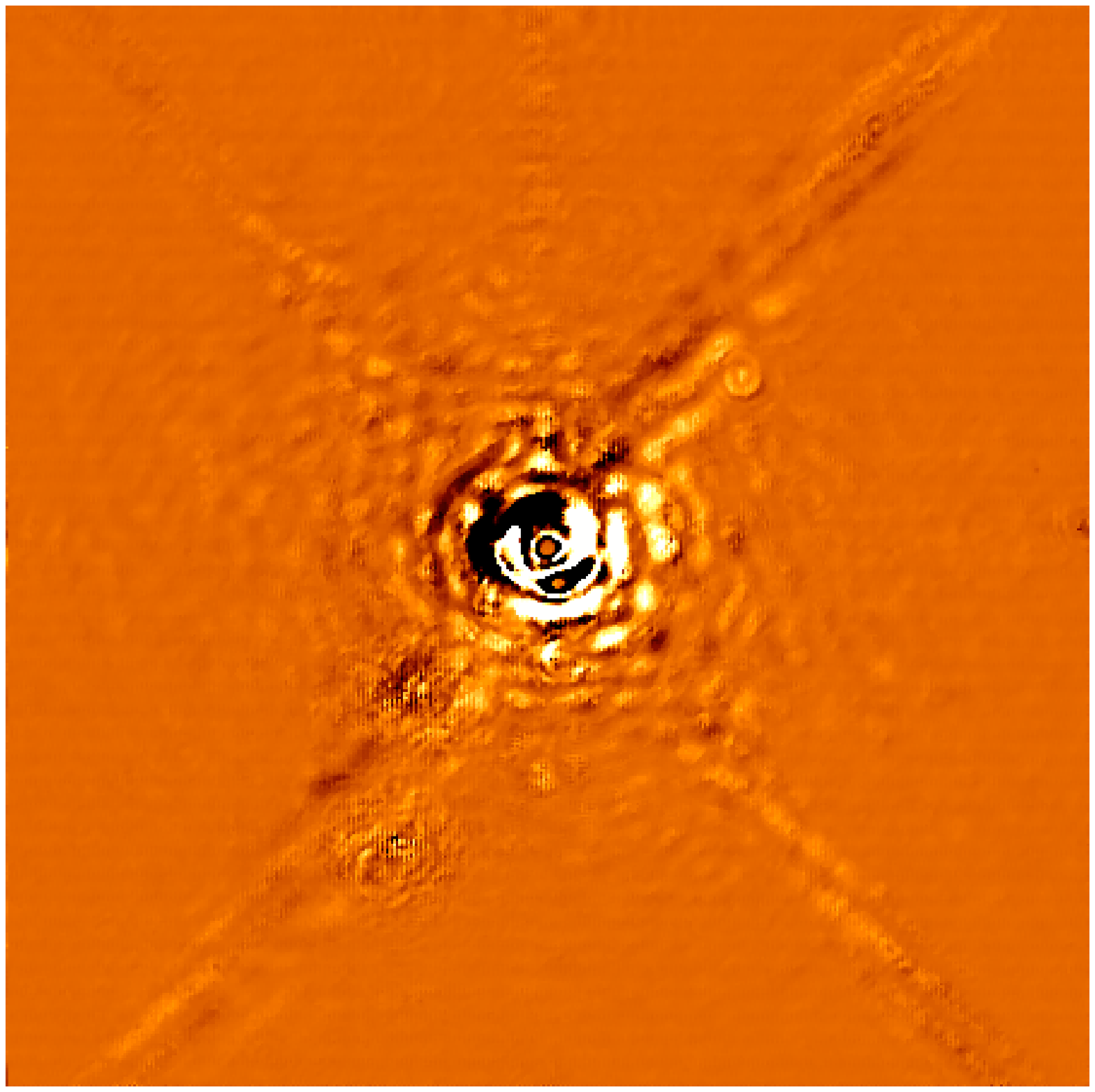}
\includegraphics[width=5.6cm]{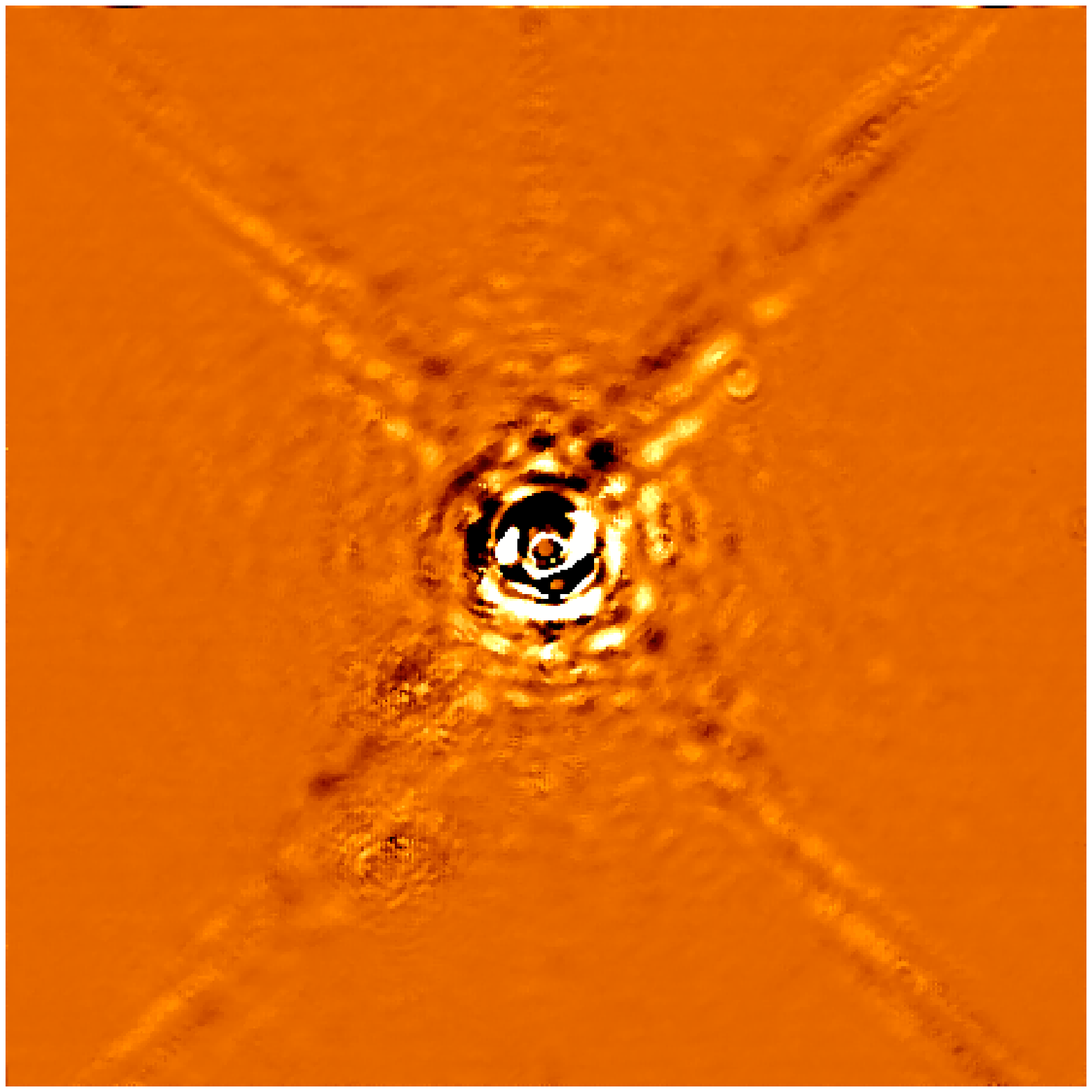}
\includegraphics[width=5.6cm]{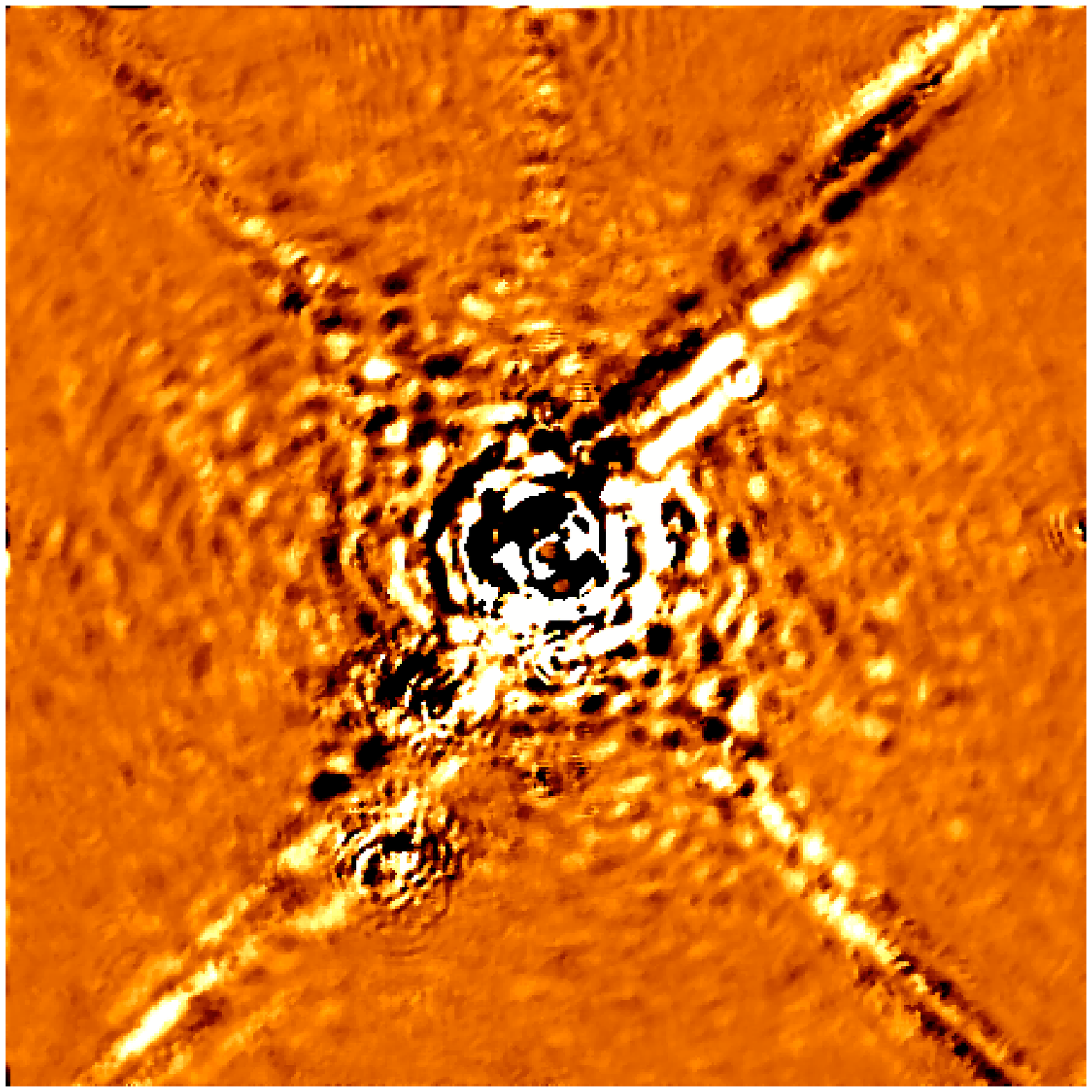}
\caption{Differential coronagraphic images. Left: difference of the $t_{0} + 3$~mn image to the reference $t_{0}$ (middle minus left image of Fig. \ref{Images}). 
Middle: difference of the $t_{0} + 6$~mn image to the reference $t_{0}$ (right minus left image of Fig. \ref{Images}). Right: difference of the $t_{0} + 165$~mn image to the reference $t_{0}$. The increase in the strength of residuals can be qualitatively observed here, and quantitatively assessed in Fig. \ref{res} as well as the scaling to the coronagraphic images.}
\label{Images2}
\end{figure*} 
\begin{figure}[htb!]
\centering
\includegraphics[width=8.5cm]{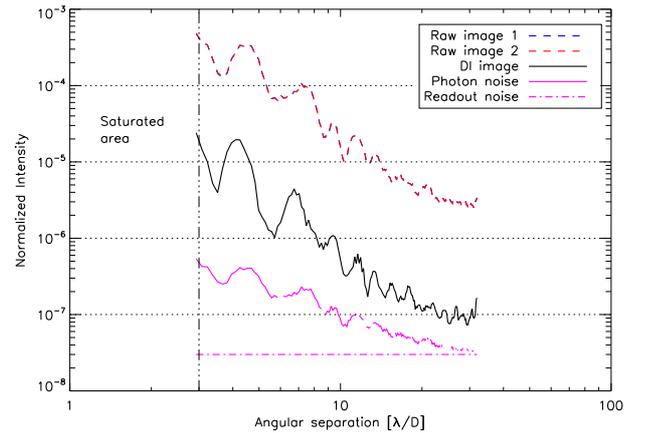}
\caption{Contrast profiles ($\mathscr{C}$) of a time series of coronagraphic images recorded at $t_0$ (dashed blue line) and $t_0 + 3$~mn (dashed red line), both are fairly identical; the detectability ($\sigma$) of the difference between both (differential image, full black line), and the associated photo noise level (pink line) and readout noise (pink dashed line).}
\label{res0}
\end{figure} 
\subsection{Data acquisition and reduction}
Raw coronagraphic images are dominated by static speckle noise. This means that the interaction between the quasi-static terms of Eq.~\ref{variance2}, being time-dependent, and static terms, assumed time-independent, can be studied through differential imaging from a time series of raw coronagraphic images.
This simply refers to the difference in intensity between an image recorded at time $t_0 + \Delta t$ and the reference image registered at $t_0$. 
The contribution from the quasi-static term is then evident from a time series of differential coronagraphic images. 
To guarantee a high speckle signal in the halo, the coronagraphic images are saturated in the image core, basically from the center to 3~$\lambda/D$.
The images are normalized to the direct, non-coronagraphic PSF flux, and taking into account a non-saturated coronagraphic image to assure an adequate calibration of the saturated images. 
\begin{figure*}[ht!]
\centering
\includegraphics[width=8.5cm]{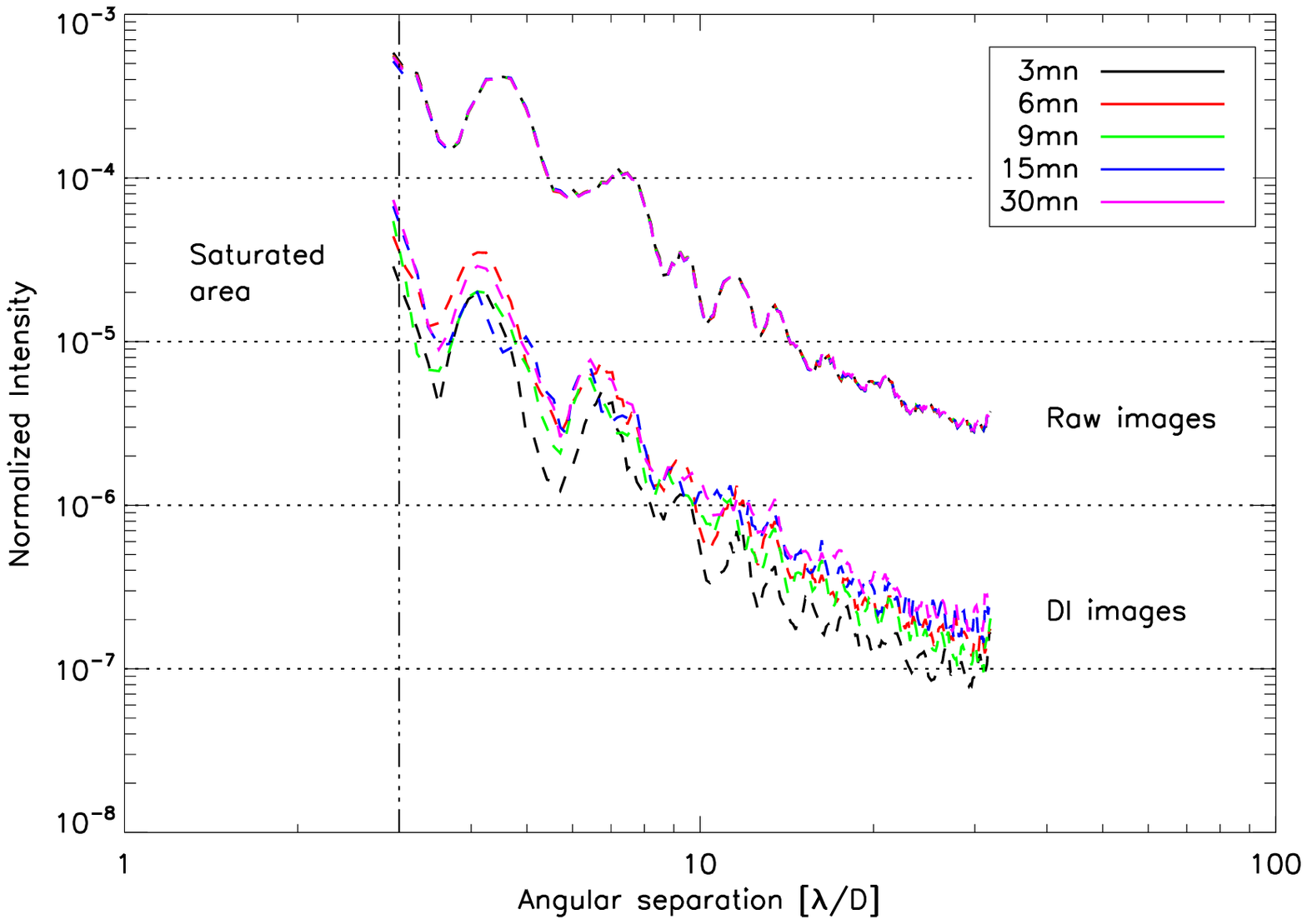}
\includegraphics[width=8.5cm]{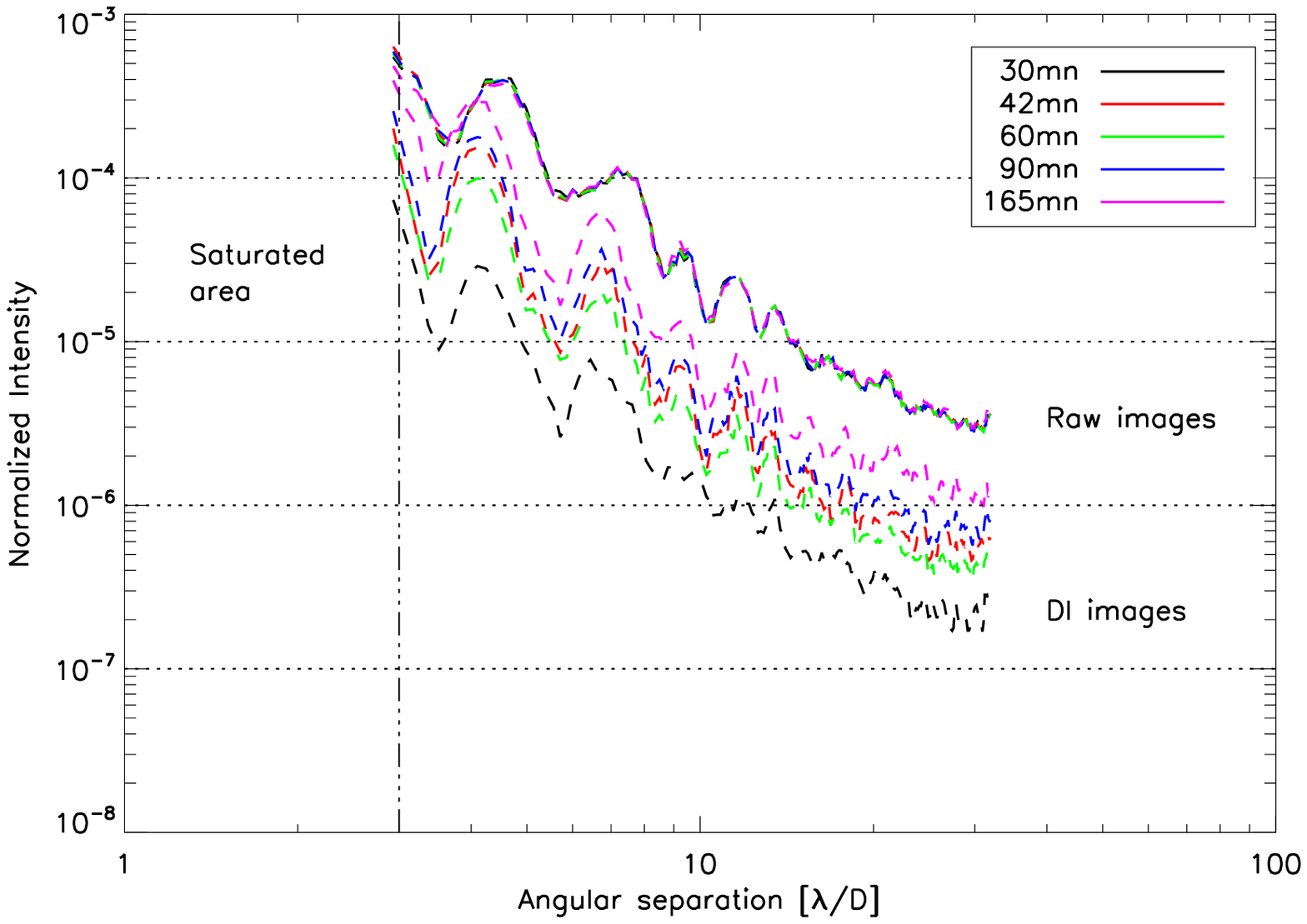}
\caption{Contrast profiles ($\mathscr{C}$) of a time series of coronagraphic images (top curves) and detectability ($\sigma$) of the differential images (subtraction of the time series of coronagraphic  images to the reference one, bottom curves). Left: timescale evolves from 3 to 30~minutes, Right: timescale evolves from 30 to 165~minutes.}
\label{res}
\end{figure*} 

Depending on the nature of the image analyzed, we applied different metrics.
The \textit{contrast} ($\mathscr{C}$) refers to the ratio of intensity in the raw coronagraphic image, averaged azimuthally at a given angular separation, to the peak intensity of the direct flux. 
When studying a differential image, implying the subtraction of 
a reference image, the average contrast is no longer suited.
The \textit{detectability} ($\sigma$) is used then, which stands for the azimuthal standard deviation measured in a ring of width $\lambda/D$. It quantifies the ability to distinguish a companion at a given angular distance. 

\noindent Each coronagraphic image of the time series corresponds to a series of 1.18~s short exposure images, averaged out over 3~mn. This was found to be the best compromise in terms of signal-to-noise level, considering inherent camera noise. 
The data-reduction process corrects for bad pixels and background, and normalizes the images with respect to exposure time and flux. The radial profile and standard deviation are then calculated from adequate regions, in particular excluding the remnant spider vane diffraction pattern. Before subtraction, a fine, sub-pixel correction of the residual Tip-Tilt component is performed on the raw images, using two ghosts present in the images as references (as seen in the left bottom corner of Fig.~\ref{Images}). These two ghosts originate from reflection in the optical train before the science arm, probably from the protecting window of the deformable mirror or the beam splitter. 
Examples of differential images are presented in Fig.~\ref{Images2}. Figure~\ref{res0} presents the contrast profiles of two raw, i.e.\ non-differential, coronagraphic images (taken at $t_0$ and $t_0 + 3$~mn in dashed lines) and the resulting detectability obtained from the difference of the two (black line). The photon noise is estimated (pink line) and guarantees that $\sigma^{2}_p$ does not limit in the noise variance expressed by Eq.~\ref{variance2}.

In Fig.~\ref{res0}, the evaluation of contrast and detectability levels are only provided for angular separation of 3~$\lambda/D$ and more, which is due to the saturation of the inner part of each image. Figure~\ref{res} presents the evolution of the contrast and detectability profiles of the raw and differential coronagraphic images evolving for 165~minutes, as obtained from a time series of images. 
These data allow studying the temporal characteristics of static and quasi-static speckles, and their interaction in a highly aberration-corrected, coronagraphic system.

\section{Analysis and interpretation}
The profiles presented in Fig.~\ref{res} clearly indicate that the detectability level degrades with time. These profiles demonstrate that raw coronagraphic images evolve temporally, being less and less correlated with the reference over time, even though such an evolution cannot be readily seen in raw images. 
Further, this degradation of the detectability is effective at all angular separations, and in a homogeneous fashion. 

\noindent As shown in Fig.~\ref{res0}, the data set is not photon noise limited, so that the contribution $\sigma^{2}_p$ from the noise variance can be neglected. In such conditions, Eq.~\ref{variance2} can be further simplified and reads:
\begin{equation}
\sigma^{2}_I \simeq \left(I^{2}_{s2} +  2 I_c I_{s2}  \right), 
\label{variance3}
\end{equation}
\noindent where the two contributing terms identify the static and quasi-static speckles. As observed in Fig.~\ref{Images} and presented in Fig.~\ref{res}, raw coronagraphic images are dominated by the static contribution, for which contrast profiles are roughly stable over time at any angular separation. Taking this into consideration, we can assume that $I_{s2}\ll I_c$. As a consequence, $I_{s2}$ can be neglected except in the cross-term, and the noise variance in the raw coronagraphic image finally becomes:
\begin{equation}
\sigma^{2}_I \simeq  2I_{c} I_{s2}. 
\label{A}
\end{equation}
This simple expression for the noise variance describes well what is observed in the data set: the degradation of the detectability level as a result of the pinning effect of the quasi-static onto the static speckles (i.e.\ an effect of interaction). This is qualitatively supported by the fact that the speckle pattern in the coronagraphic image is stable over time.  

A breakdown of this pinning effect can be carried out at the level of the differential images. A similar expression of the noise variance for the differential images ($\sigma_{DI}$) can be derived as the difference of Eq.~\ref{A} evaluated at $t_0 + \Delta t$, to that of the reference, at $t_0$. It reads: 
\begin{equation}
\sigma^{2}_{DI} \simeq  2I_{c}\Delta I_{s2},  
\label{B}
\end{equation}
\noindent where $\Delta I_{s2}$ represents the quasi-static evolution between the two successive images. Therefore, the quasi-static contribution can be expressed as:
\begin{equation}
\Delta I_{s2} \simeq \frac{\sigma^{2}_{DI}}{2I_{c}}.
\label{C}
\end{equation}

\noindent A general expression of the speckle intensity \citep{Racine99} is:
\begin{equation}
I_{speckle} \approx \frac{(1 - S)}{0.34},
\label{D}
\end{equation}
\noindent where S can be related to the wavefront error $\phi$ using Mar\'{e}chal's approximation \citep{BW}: $S \approx 1 - \left( \frac{2\pi \phi}{\lambda}\right)^{2}$. With this, and using Eqs.~\ref{C} and \ref{D}, the contribution from quasi-static speckles to the wavefront error in the differential images finds an approximated expression in:
\begin{equation}
\Delta \phi_{s2} \simeq \frac{\lambda}{2\pi \sqrt6} \times \frac{\sigma_{DI}}{\sqrt I_{c}}.
\label{E}
\end{equation}
\noindent Similarly, the wavefront error associated to the static contribution (assuming that $I_{s2}\ll I_c$) can be expressed as:
\begin{equation}
\phi_{c} \simeq \frac{\lambda }{\pi \sqrt6} \times \sqrt{I_c}.
\label{F}
\end{equation}
Hence, using Eqs.~\ref{E} and \ref{F}, the analysis of both direct and differential images becomes possible. The static wavefront error amplitude has been evaluated for the raw coronagraphic images using Eq.~\ref{F}, and for our data set converges to the value of 48~nm~rms ($\lambda=1.65$~$\mu$m). This is in fair agreement with our initial expectation (50 to 75~nm~rms). 



\begin{figure}[ht!]
\centering
\includegraphics[width=8.5cm]{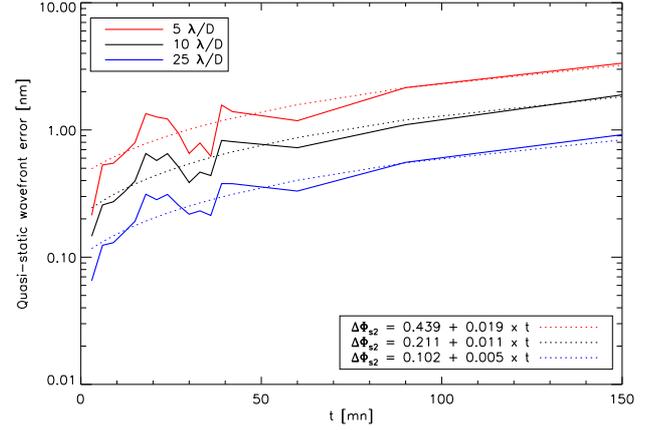}
\caption{Time variability of wavefront error due to quasi-static speckles, evaluated at various angular separations (observational data).}
\label{res2}
\end{figure}

From our time series of differential images, using Eq.~\ref{E}, we derive the quasi-static wavefront error contribution \textit{per Fourier component}, of the pinning effect at several angular separations. 
Figure~\ref{res2} shows the temporal evolution of $\phi_{s2}$ at 5, 10, and 25~$\lambda/D$, i.e.\ the quasi-static wavefront error (rms) as function of time at several angular separations. It clearly indicates that  $\phi_{s2}$ is time-dependent and increases with time, justifying the constant degradation of detectability observed in Fig.~\ref{res}. This is true for all angular separations, and the shorter the separation the higher the amplitude. 

The power law of the temporal evolution of the quasi-static wavefront error at the three angular separations is derived and exhibits a similar tendency. It can be fitted by a linear function of time. The slope of the linear fit decreases with the angular separation, i.e.\ at 5~$\lambda/D$ the increase of the quasi-static wavefront error with time is roughly two times faster than at 10~$\lambda/D$. The parameters for the linear fits are presented in the legend of  Fig.~\ref{res2}, and can be used to extrapolate a model for quasi-static speckle evolution at least under laboratory conditions. From these linear fits derived for the data at 5, 10, and 25~$\lambda/D$, and considering that the difference in amplitude is not significant (below 1 nm rms over an hour), we attempt to generalize the expression of the power law for any angular position in the field, which reads as the following approximation:
\begin{equation}
\Delta \phi_{s2} \simeq 0.250 + 0.012 \times t,
\label{FIT}
\end{equation}
\noindent where $t$ is the time in minutes, and $\Delta \phi_{s2}$ is expressed in nm~rms. Though nothing supports the universality of the parameters for the fit, 
the linear evaluation with time is clearly established in a laboratory environment. 
\begin{figure*}[ht!]
\centering
\includegraphics[width=8.5cm]{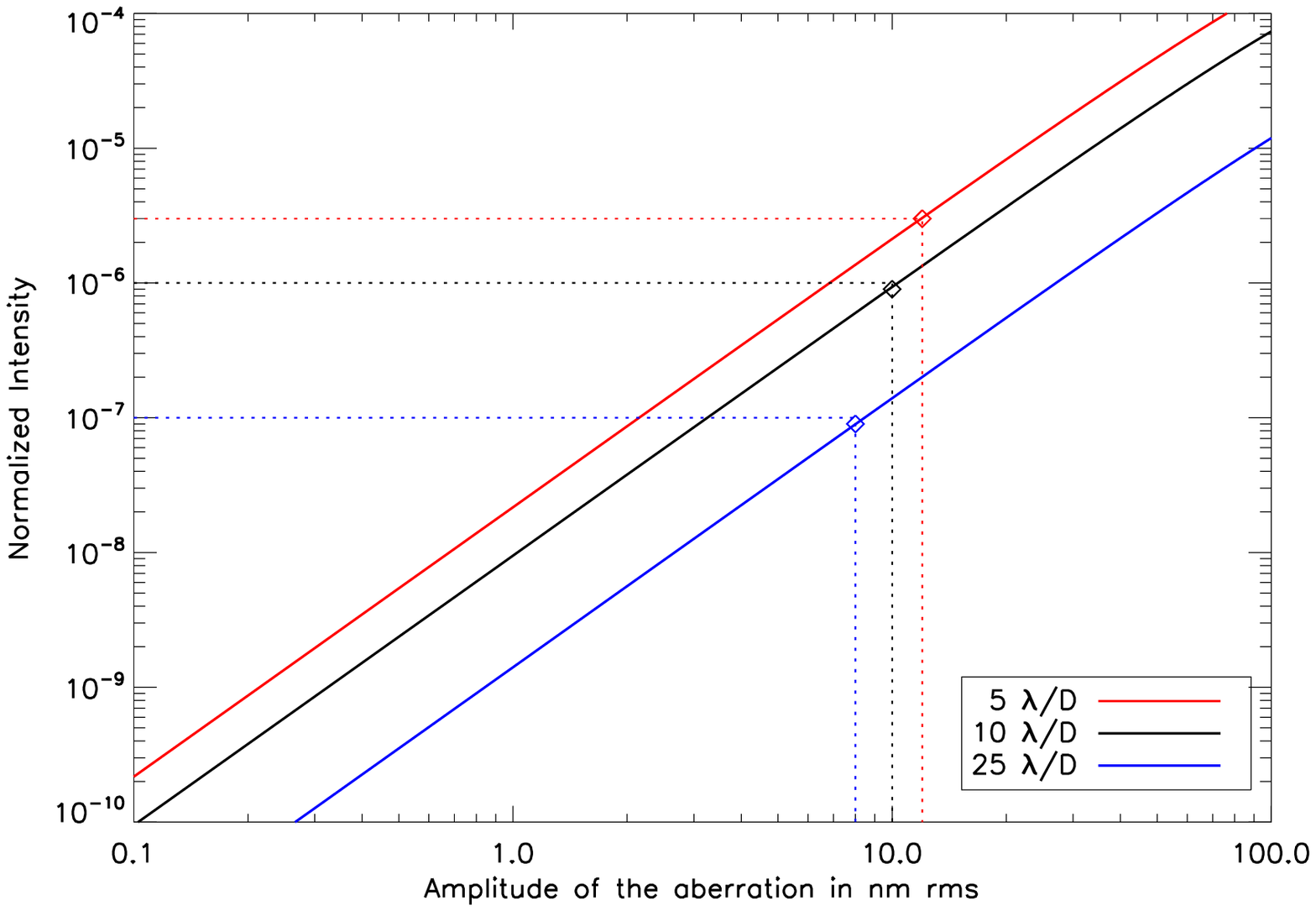}
\includegraphics[width=8.5cm]{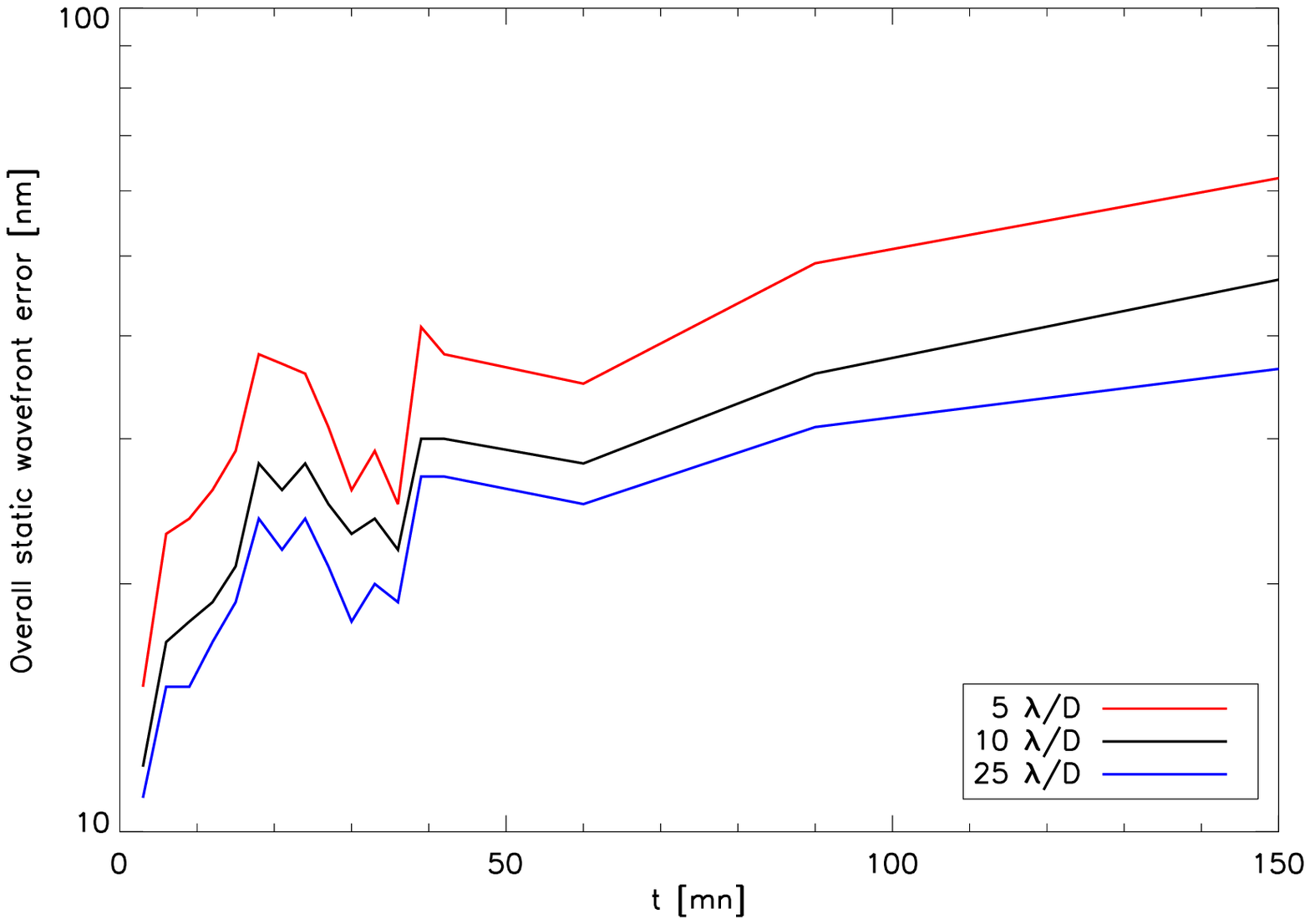}
\caption{Left: Detectability ($\sigma$) of the differential image at various angular separations, with increasing difference in static aberration between the reference and the second image (simulated data). 
Right: time variability of wavefront error due to static speckles, estimated at various separations (simulated data, assuming $f^{-2}$ for the power law).}
\label{res1}
\end{figure*} 
The linear temporal dependency was expected since the derivative of Eq. \ref{FIT} must be a constant to ensure that the wavefront degradation over $\Delta t$ is independent on the actual value of the reference image taken arbitrary here at $t_0$. Higher order terms would therefore make no physical sense. Only a linear relation with time is consistent with this condition.  

Quantitatively, the level of quasi-static wavefront error responsible for the loss of contrast in the differential images (Fig.~\ref{res}) is found to be in the regime of 1~$\AA$ to 1~nm over an hour of time. In addition, for all curves (Fig.~\ref{res2}), a characteristic knee between 30 and 40~minutes is observed, during which the contrast degrades significantly (Fig.~\ref{res}), though its origin is not well understood. Nevertheless, the jump between 30 and 40~minutes does not deviate far from the statistics inherent to the data, and is likely a result of a disturbance during the measurement series.  
To assess whether there is an actual decorrelation over this specific timescale, we re-analyzed the data assuming the $t_0 + 12$~mn as the reference image. In such condition, the knee is still observable but at a timescale of 20 to 30~minutes in this case. This simple test clearly indicates that the observed knee in the data is due to a disturbance during the measurement rather than a statistical event.

Additionally, with simulations based on an IDL Fraunhofer propagation code which reflects our test setup and experimental conditions, we estimate the level of amplification of the static contribution (i.e.\ the total wavefront error) that can explain the detectability level obtained from our differential image data between $t_0$ and $t_0 + 3$~mn (black curve, Fig.~\ref{res0}, similarly in the left figure of Fig.~\ref{res}).
The simulation assumes as-manufactured coronagraphic components and a typical power law of manufactured optics (PSDs with $f^{-2}$ variation, e.g.\ \citet{D02}, where $f$ is the spatial frequency), to generate two coronagraphic images based on the same static aberration pattern but with slight different amplitude levels. This simulates the effect resulting from the interaction with pinned speckles, which we examine here. 
We found that $\sim$~10~nm~rms overall difference in static wavefront error between two images converges to similar contrast levels as the ones observed in our data (Fig.~\ref{res1}, left). The detectability according to simulated and observed data has been matched at various angular separations to assess the error resulting from our estimation 
($\pm$~2~nm in this case, see Fig.~\ref{res1}, left). 

This result is supported by the rough estimate that can be derived from Eq.~\ref{C} using similar values, which for the contribution from quasi-static speckles converges to a nanometer level. In addition, the overall static wavefront error can be estimated similarly for all our data time gaps, and its temporal evolution can be deduced in a similar way as done previously for the quasi-static contribution. The total static wavefront error as a function of time is presented in Fig.~\ref{res1} (right). Again, to assess the error from the estimate, it has been derived at various angular separations. It is observed that the estimates derived from various angular separations are less and less compatible over time. A fair agreement is found on a short time scale, while the estimates differ significantly on an intermediate and longer time scale. However, the power law derived for the temporal evolution of the total wavefront error due to static aberrations is found to be similar at all angular separations and also to be linear, as it was observed for the quasi-static speckles. 

In conclusion, the amplification of the 50~nm~rms static aberrations present in the system by $\gtrapprox$~10~nm~rms can be explained by the effect of pinned, quasi-static speckles at the level of few angstroms to nanometers. This effect is assumed to reflect thermal and mechanical instabilities of the optical bench.  

\section{Discussion}
The temporal evolution of quasi-static speckles is a key parameter to predict the performance of upcoming XAO imagers. When the noise is dominated by quasi-static speckles, it still can be calibrated but to a certain extent.
Using a reference image quasi-static noise can be subtracted, but its temporal evolution ultimately limit this possibility, and in particular it is understood that some timescales have a larger impact than others.
For instance, a typical hour-long ADI observing sequence provides a partial self-calibration of the residuals after a rotation of $\sim1\lambda/D$ at a given angular separation, which generally requires less than few minutes (e.g., 5 to 7 mn at 1$\arcsec$ on a 8-m class telescope for stars near the meridian in $H$-band), though it depends on wavelength. Residual speckles with decorrelation times faster than the time needed to obtain the ADI reference image cannot then be removed, while quasi-static speckles associated with larger timescales can largely be subtracted.
In this context, the observed decorrelation time in our data ($\ll$3~mn integration time) being slightly dependent on field position, provides relevant information to understand and predict the limitation of existing and upcoming high-contrast instruments.
In addition, it is remarkable to see that the decorrelation timescales observed in our data at very high Strehl ratio and the ones observed by \citet{Marois06} for long sequence of ADI at low $\sim$20$\%$ Strehl ratio (see Fig. 2 and 4 of that paper) are of the same order (few minutes).

Nonetheless, it should be pointed out that our results have been obtained in laboratory environment and are thus valid in such conditions unless demonstrated otherwise, and scaling them to real instrument is highly non-trivial. 
Indeed, nothing supports the universality of the quasi-static speckle decorrelation time observed in our data, which is significantly dependent of the instrument environment (temperature or pressure changes, mechanical flexures, etc...). 
Furthermore, as initially discussed in \citet{Martinez10}, our contrast levels must be carefully consider as direct comparison with what real high-Strehl instruments will provide is neither easy.
Upcoming XAO imagers are much more complicated than our experimental demonstrator, and will make the use of several calibration procedures such as use of a reference star, from the observation of the same image but obtained at different orientations (ADI), wavelengths (dual-beam imaging), or polarizations (differential polarimetry), or calibration image though an Integral Field Spectrograph. 

\section{Conclusion}
The most significant result of this analysis of the static and quasi-static aberrations observed in a time series of extreme adaptive optics-corrected coronagraphic images, is that they exhibit a linear power law with time and are interacting through the pinning effect. 
We examine and discuss this effect using the statistical model of the noise variance in high-contrast imaging proposed by \citet{Soummer07}. 
The effect of pinning quasi-static to static speckles as described by the expression for the variance (Eq.~\ref{variance2}) is found to reflect the situation in our data set, which was obtained in a laboratory environment, fairly well.

We found that quasi-static speckles, fast-evolving on the level of a few angstroms to nanometers, explain the evolution of our differential imaging contrast in the halo from $10^{-6}$ to $10^{-7}$ through amplification of the systematics. It is believed that this effect is a consequence of thermal and mechanical instabilities of the optical bench. The HOT bench is localized in a classical laboratory setting, where the enclosure of the bench is not thermally actively controlled, but the room temperature was monitored, and found stable at the level of few tenths of degree.
This result emphasizes the importance of such stability for the next generation of high-contrast instruments. In addition, for a further increase in the dynamic range of instruments, reducing the level of static speckles ($I_c$) through speckle reduction techniques, e.g.\ speckle nulling, is shown to be critically important.

The temporal evolution of quasi-static speckles is required for a quantification of the gain expected when using angular differential imaging, and to determine the interval on which speckle nulling techniques must be carried out. 
In our data, quasi-static speckles are observed to be temporally instable over a timescale of a few seconds 
($\ll$3~mn integration time) and demonstrate a linear power law with time. 
Although nothing supports the universality of this estimate, in order to guarantee the high contrast level required to detect faint companions, non-common path wavefront errors must be sensed and corrected at a timescale faster than the typical speckle decorrelation time.  

\acknowledgements
The activity outlined in this paper has been funded as part of the European Commission, Seventh Framework Programme (FP7), Capacities Specific Programme, Research Infrastructures; specifically the FP7, Preparing for the construction of the European Extremely Large Telescope Grant Agreement, contract number INFRA-2007-2.2.1.28.

\newpage
\bibliography{MyBiblio} 

\begin{thebibliography}{22}
\expandafter\ifx\csname natexlab\endcsname\relax\def\natexlab#1{#1}\fi

\bibitem[{{Aller-Carpentier} {et~al.}(2008){Aller-Carpentier}, {Kasper},
  {Martinez}, {Vernet}, {Fedrigo}, {Soenke}, {Tordo}, {Hubin}, {Verinaud},
  {Esposito}, {Pinna}, {Puglisi}, {Tozzi}, {Quiros}, {Basden}, {Goodsell},
  {Love}, \& {Myers}}]{HOTbench}
{Aller-Carpentier}, E., {Kasper}, M., {Martinez}, P., {et~al.} 2008, in Society
  of Photo-Optical Instrumentation Engineers (SPIE) Conference Series, Vol.
  7015

\bibitem[{{Beuzit} {et~al.}(2008){Beuzit}, {Feldt}, {Dohlen}, {Mouillet},
  {Puget}, {Wildi}, {Abe}, {Antichi}, {Baruffolo}, {Baudoz}, {Boccaletti},
  {Carbillet}, {Charton}, {Claudi}, {Downing}, {Fabron}, {Feautrier},
  {Fedrigo}, {Fusco}, {Gach}, {Gratton}, {Henning}, {Hubin}, {Joos}, {Kasper},
  {Langlois}, {Lenzen}, {Moutou}, {Pavlov}, {Petit}, {Pragt}, {Rabou}, {Rigal},
  {Roelfsema}, {Rousset}, {Saisse}, {Schmid}, {Stadler}, {Thalmann}, {Turatto},
  {Udry}, {Vakili}, \& {Waters}}]{SPHERE}
{Beuzit}, J.-L., {Feldt}, M., {Dohlen}, K., {et~al.} 2008, in Society of
  Photo-Optical Instrumentation Engineers (SPIE) Conference Series, Vol. 7014

\bibitem[{{Beuzit} {et~al.}(1997){Beuzit}, {Mouillet}, {Lagrange}, \&
  {Paufique}}]{Beuzit97}
{Beuzit}, J.-L., {Mouillet}, D., {Lagrange}, A.-M., \& {Paufique}, J. 1997,
  \aaps, 125, 175

\bibitem[{{Boccaletti} {et~al.}(2003){Boccaletti}, {Chauvin}, {Lagrange}, \&
  {Marchis}}]{Boccaletti03}
{Boccaletti}, A., {Chauvin}, G., {Lagrange}, A.-M., \& {Marchis}, F. 2003,
  \aap, 410, 283

\bibitem[{{Boccaletti} {et~al.}(2004){Boccaletti}, {Riaud}, {Baudoz},
  {Baudrand}, {Rouan}, {Gratadour}, {Lacombe}, \& {Lagrange}}]{Boccaletti04}
{Boccaletti}, A., {Riaud}, P., {Baudoz}, P., {et~al.} 2004, \pasp, 116, 1061

\bibitem[{{Born} \& {Wolf}(1993)}]{BW}
{Born}, M. \& {Wolf}, E. 1993, Cambridge University Press

\bibitem[{{Duparr{\'e}} {et~al.}(2002){Duparr{\'e}}, {Ferre-Borrull}, {Gliech},
  {Notni}, {Steinert}, \& {Bennett}}]{D02}
{Duparr{\'e}}, A., {Ferre-Borrull}, J., {Gliech}, S., {et~al.} 2002, \ao, 41,
  154

\bibitem[{{Hinkley} {et~al.}(2007){Hinkley}, {Oppenheimer}, {Soummer},
  {Sivaramakrishnan}, {Roberts}, {Kuhn}, {Makidon}, {Perrin}, {Lloyd},
  {Kratter}, \& {Brenner}}]{Hinkley07}
{Hinkley}, S., {Oppenheimer}, B.~R., {Soummer}, R., {et~al.} 2007, \apj, 654,
  633

\bibitem[{{Hinkley} {et~al.}(2011){Hinkley}, {Oppenheimer}, {Zimmerman},
  {Brenner}, {Parry}, {Crepp}, {Vasisht}, {Ligon}, {King}, {Soummer},
  {Sivaramakrishnan}, {Beichman}, {Shao}, {Roberts}, {Bouchez}, {Dekany},
  {Pueyo}, {Roberts}, {Lockhart}, {Zhai}, {Shelton}, \& {Burruss}}]{P1640}
{Hinkley}, S., {Oppenheimer}, B.~R., {Zimmerman}, N., {et~al.} 2011, \pasp,
  123, 74

\bibitem[{{Hodapp} {et~al.}(2008){Hodapp}, {Suzuki}, {Tamura}, {Abe}, {Suto},
  {Kandori}, {Morino}, {Nishimura}, {Takami}, {Guyon}, {Jacobson},
  {Stahlberger}, {Yamada}, {Shelton}, {Hashimoto}, {Tavrov}, {Nishikawa},
  {Ukita}, {Izumiura}, {Hayashi}, {Nakajima}, {Yamada}, \& {Usuda}}]{HiCIAO}
{Hodapp}, K.~W., {Suzuki}, R., {Tamura}, M., {et~al.} 2008, in Society of
  Photo-Optical Instrumentation Engineers (SPIE) Conference Series, Vol. 7014

\bibitem[{{Lafreni{\`e}re} {et~al.}(2007){Lafreni{\`e}re}, {Marois}, {Doyon},
  {Nadeau}, \& {Artigau}}]{2007ApJ...660..770L}
{Lafreni{\`e}re}, D., {Marois}, C., {Doyon}, R., {Nadeau}, D., \& {Artigau},
  {\'E}. 2007, \apj, 660, 770

\bibitem[{{Macintosh} {et~al.}(2008){Macintosh}, {Graham}, {Palmer}, {Doyon},
  {Dunn}, {Gavel}, {Larkin}, {Oppenheimer}, {Saddlemyer}, {Sivaramakrishnan},
  {Wallace}, {Bauman}, {Erickson}, {Marois}, {Poyneer}, \& {Soummer}}]{GPI}
{Macintosh}, B.~A., {Graham}, J.~R., {Palmer}, D.~W., {et~al.} 2008, in Society
  of Photo-Optical Instrumentation Engineers (SPIE) Conference Series, Vol.
  7015, Society of Photo-Optical Instrumentation Engineers (SPIE) Conference
  Series

\bibitem[{{Marois} {et~al.}(2006{\natexlab{a}}){Marois}, {Lafreni{\`e}re},
  {Doyon}, {Macintosh}, \& {Nadeau}}]{Marois2006}
{Marois}, C., {Lafreni{\`e}re}, D., {Doyon}, R., {Macintosh}, B., \& {Nadeau},
  D. 2006{\natexlab{a}}, \apj, 641, 556

\bibitem[{{Marois} {et~al.}(2003){Marois}, {Nadeau}, {Doyon}, {Racine}, \&
  {Walker}}]{Marois03}
{Marois}, C., {Nadeau}, D., {Doyon}, R., {Racine}, R., \& {Walker}, G.~A.~H.
  2003, in IAU Symposium, Vol. 211, Brown Dwarfs, ed. {E.~Mart{\'{\i}}n},
  275--+

\bibitem[{{Marois} {et~al.}(2006{\natexlab{b}}){Marois}, {Phillion}, \&
  {Macintosh}}]{Marois06}
{Marois}, C., {Phillion}, D.~W., \& {Macintosh}, B. 2006{\natexlab{b}}, in
  Society of Photo-Optical Instrumentation Engineers (SPIE) Conference Series,
  Vol. 6269, Society of Photo-Optical Instrumentation Engineers (SPIE)
  Conference Series

\bibitem[{{Martinez} {et~al.}(2011){Martinez}, {Aller-Carpentier}, {Kasper},
  {Boccaletti}, {Dorrer}, \& {Baudrand}}]{Martinez11}
{Martinez}, P., {Aller-Carpentier}, E., {Kasper}, M., {et~al.} 2011, \mnras,
  414, 2112

\bibitem[{{Martinez} {et~al.}(2010){Martinez}, {Carpentier}, \&
  {Kasper}}]{Martinez10}
{Martinez}, P., {Carpentier}, E.~A., \& {Kasper}, M. 2010, \pasp, 122, 916

\bibitem[{{Martinez} {et~al.}(2009){Martinez}, {Dorrer}, {Aller Carpentier},
  {Kasper}, {Boccaletti}, {Dohlen}, \& {Yaitskova}}]{microdots1}
{Martinez}, P., {Dorrer}, C., {Aller Carpentier}, E., {et~al.} 2009, \aap, 495,
  363

\bibitem[{{Oppenheimer} {et~al.}(2001){Oppenheimer}, {Golimowski}, {Kulkarni},
  {Matthews}, {Nakajima}, {Creech-Eakman}, \& {Durrance}}]{Oppenheimer01}
{Oppenheimer}, B.~R., {Golimowski}, D.~A., {Kulkarni}, S.~R., {et~al.} 2001,
  \aj, 121, 2189

\bibitem[{{Racine} {et~al.}(1999){Racine}, {Walker}, {Nadeau}, {Doyon}, \&
  {Marois}}]{Racine99}
{Racine}, R., {Walker}, G.~A.~H., {Nadeau}, D., {Doyon}, R., \& {Marois}, C.
  1999, \pasp, 111, 587

\bibitem[{{Soummer} {et~al.}(2003){Soummer}, {Aime}, \& {Falloon}}]{Soummer03}
{Soummer}, R., {Aime}, C., \& {Falloon}, P.~E. 2003, \aap, 397, 1161

\bibitem[{{Soummer} {et~al.}(2007){Soummer}, {Ferrari}, {Aime}, \&
  {Jolissaint}}]{Soummer07}
{Soummer}, R., {Ferrari}, A., {Aime}, C., \& {Jolissaint}, L. 2007, \apj, 669,
  642

\end{thebibliography}
\end{document}